# PIXEL: Interactive Light System Design Based On Simple Gesture Recognition

ZOU Xuedan          Huan University, Changsha, China

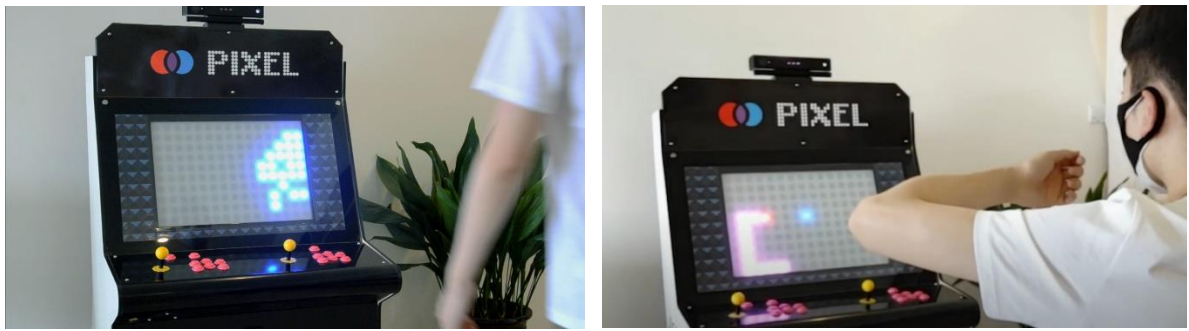

Figure 1: The outlook of our system. Left: Pixel creature follows user. Right: User plays SNAKE game.

## ABSTRACT

In this project, by utilizing the real-time human gestures captured by Kinect, we attempted to provide a self-made interactive light system PIXEL for interacting with the visitor to play a simple SNAKE game. By paralleling the low power single color LED lights and lighting up them separately or together, we provided a big LED board with multiple colors while at the time safe enough without using high tension electricity. We analysed the factors that influence the final visual effect of the light pixel, the novel gestures that can be recognized well by current computer vision algorithms and did several experiments to decide the final interactive method. We also believe this project provides a way to help people reconsider the relationship between the old and the new and the possibility to bring old things reborn by the new technologies.

## INTRODUCTION

With the development of computer vision, computers can now recognize many human behaviors accurately, which provides more possibilities for real-time no-touched interactive systems. At the same time, these fast developing technologies greatly change our digital lives and make many old things fade away. Considering the relationship between the fast growing new technologies and the fading old digital culture, we aim to purpose an interactive system using those new technologies to interactive with the user the past fading culture, thus to discuss this confliction: whether they can be in fact an organic whole.

To achieve this goal, we have to build a system being used as the present hardware and design the content to interact. We focus our theme on the digital culture since it is a wonderful carrier, which was first defined in 1982 [1], of the modern computer vision and computer graphics technology. Compared with using a normal screen to display our content, we chose a more artistic way to express the pixel feeling by building a pixel based LED light matrix. In fact there are several previous researches on the pixel based interactive display system. One typical type is a dynamic actuation of digital contents with physical objects such as square pins [2,3,4]. However, the present of visual images of those systems always need an additional projector and thus the visual appearance can be greatly messed up in public places. As a result, we chose the pixel light system instead. Besides, compared with most huge light matrix used in museums need a high power supply and can be extremely dangerous to operate, we purpose a light matrix system with only low power supply, which can be used safely at home. We finally chose some classic pixel characters from video games like Space Invaders, a modified Game of Life [5] and a classic SNAKE game to interact with the user.

## SYSTEM DESIGN

In this system, there is a Kinect placed on the Arcade. It can collect the person's real-time skeleton data (with its depth information) and tell the laptop. The laptop then computes the image output result through the codes and give the feedback to the microprocessor, which controls on and off of each single LED. Finally the person notice the interactive result and do the next step, the process iterates again.

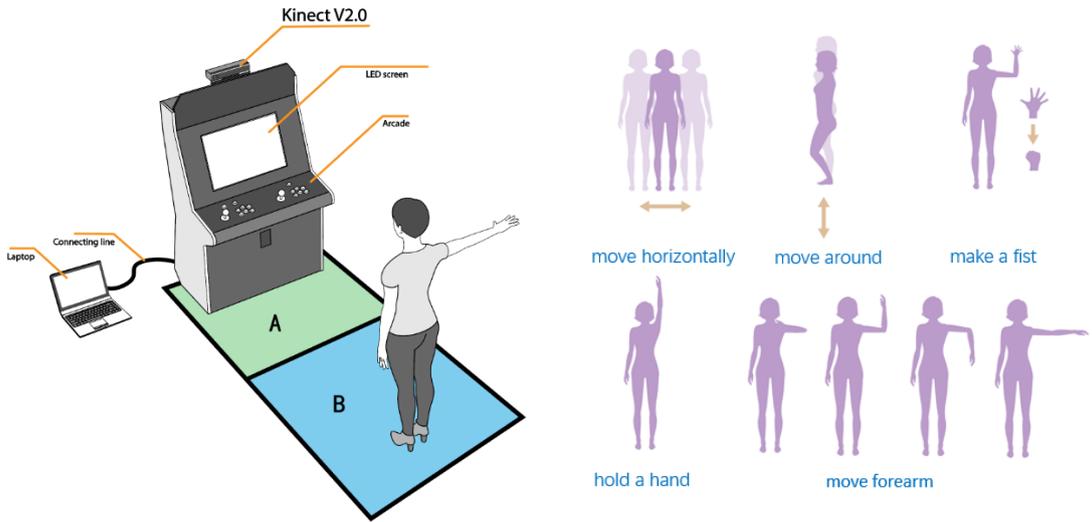

Figure 2: Left: The system map of our work.   Right: The gestures chosen that can be recognized.

Since we wanted to make sure the accuracy rating of recognizing, we found out that Kinect can work well in the distance between 0.5 to 4.5 meter. [6,7] According to the distance between the person and the Arcade, the space is divided into two areas, A and B. In area A, people can play games and in area B there are some random animes which can be interacted with people. The depth data of the real-time skeleton data can tell the system whether the person is in area A or area B. Some simple but natural human gestures are chosen to interact and the following image shows all of them.

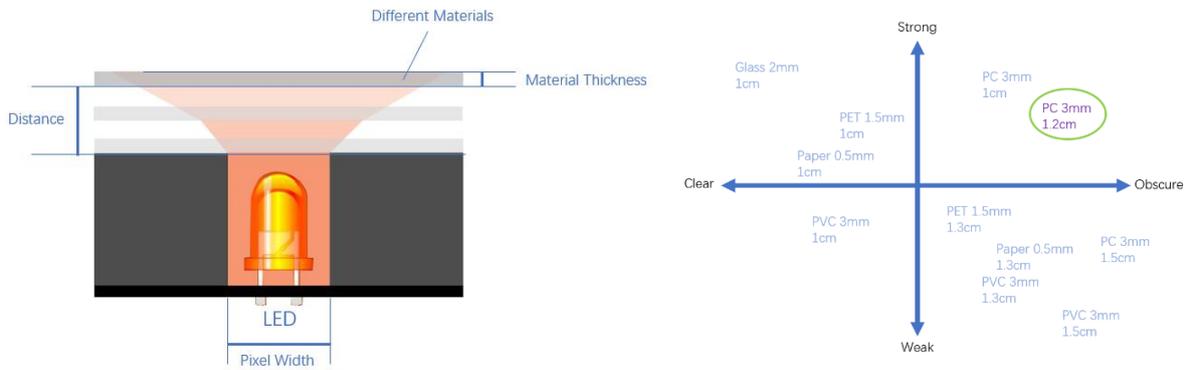

Figure 3: Left: The factors that decide the final visual effect of our system.    Right: We measured the visual effect produced by different materials with different thickness and in different distances by dividing them into strong or weak and clear or obscure. .

For the consideration of safety, instead of using high power strong light LED, we chose the low power LED and by paralleling some of them together and let the light passes through some specific materials placed beyond to provide pixel light well. The principle of producing those beautiful pixel lights is shown in the following image. The visual effect is strongly connected with the pixel (hole) width, the chosen light-passing material, the sickness of this material and the distance between the material and the LED light. we have tested multiple possible materials in multiple distances to get the best visual result.

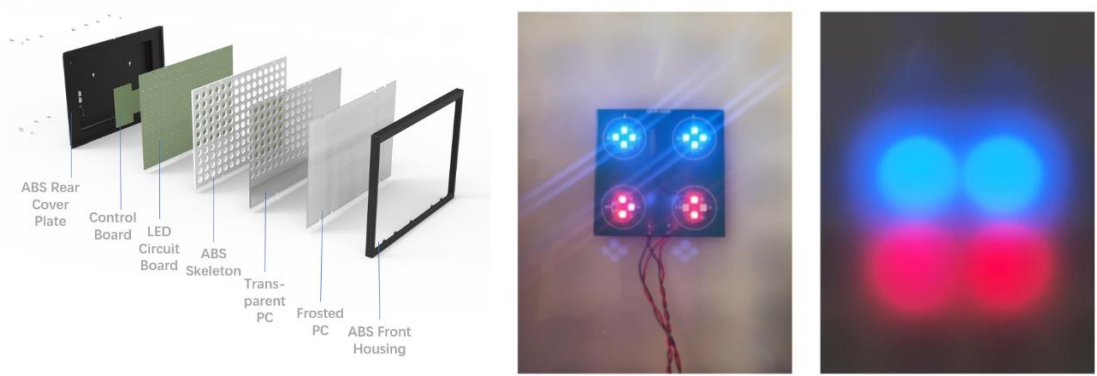

Figure 4: Left: The structure of our display system.    Right: A prototype of the LED light we used and the visual effect after it passes through the PC material.

Here in order to low down the complexity of the circuit design, we chose a red and a blue LED light, both in single color. When the blue and red LED light turn on together we can see the purple color, which is an obvious result through the principle of Color Mixing. So by this mean we can control in total 4 status of each LED light (red, blue, purple and off) and use them to get some meaningful images. All of these circuits are put all together in the designed hardware. The control board gets the information from the laptop and controls the LED circuit board. Lights from small LED lights pass through the ABS skeleton to get together, forming the pixel light. Finally the pixel light passes through the transparent PC to get the final visual result.

## CONTENTS AND MEANING

Since the total size of pixels is quite small(only 15x10), it is a great challenge to give a meaningful output. Inspired by the Pixel Culture that we have discussed before in the Introduction section, we decided to present something in the past. So we chose a classic SNAKE game, some pixel characters in some games, and a modified GAME OF LIFE to present.

In general, our interactive journey goes as the following. In the beginning the pixel creature on the screen will follow the user. And if the user holds up his hand and turns the hand from open to close then the stuffs on the screen will be changed. The screen shows a random pixel creature or a GAME OF LIFE anime. Come close enough and the system will jump into the SNAKE Game mode. First the user holds up one side of hand (left or right) then the direction of forearm will lead the snake to move. The velocity of the snake will increase gradually as the score goes higher. After the player dies in the game, the final score will scroll on the screen and the journey begins from the start.

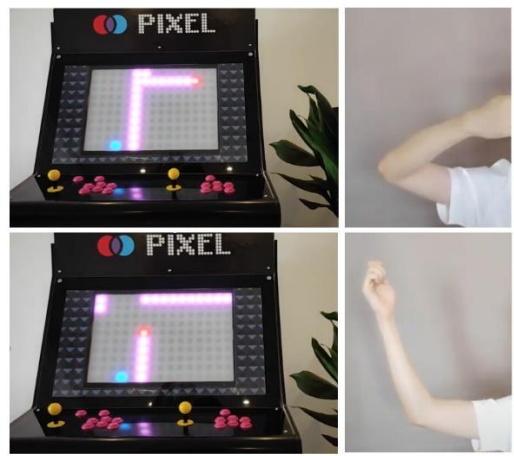

```
GAME OF LIFE GENERAL CODE

no_colored_pixel=true;
for each pixel_not_border in new_img
If 8_neighbor_colored in old_img==3
    pixel_not_border.colored()
    no_colored_pixel=false
else
    pixel_not_border.uncolored()
for random(n) times
    random_border_pixel.uncolored() in new_img
If no_colored_pixel==true
for random(m) times
    pixel_not_border.colored() in new_img
```

Figure 5: Left: The pseudo-code of our GAME OF LIFE.   Right: The user uses the forearm direction to lead the movement of the snake.

We have invited 30 players, who are aged from 17 to 45 and are randomly chosen people without any mental or physical diseases, to play our system PIXEL and asked them about their feeling after playing it. Most players got confused at first about the way to interact but then soon got familiar with the system after being demonstrated. Almost all of the players claimed that they were impressed and felt totally different with the traditional way playing SNAKE game by physical touched devices. We made success to make them reconsider the relationship between old and new and some convinced that they felt the consistent humanism thoughts under the fast growing technologies.

## CONCLUSION

The presented system is work-in-progress. We proposed PIXEL, an interactive light system that handles the Kinect captured gestures data and presents contents related to pixel culture. We believe our system can bring people new feelings on past familiar contents and can inspire people's thoughts on the relationship between the past and the future. Further more the system can be broaden to other applications like helping the old doing physical exercise, helping the physical injured patients to get recovery. Many potential situations need to be discussed in the future.